# Time-dependent linear response of an inhomogeneous Bose superfluid: Microscopic theory and connection to current-density functional theory

M. L. Chiofalo, A. Minguzzi* and M. P. Tosi

*Istituto Nazionale di Fisica della Materia and Classe di Scienze,*

*Scuola Normale Superiore, I-56126 Pisa, Italy*

**Abstract** - The dynamics of a confined fluid of Bose atoms is treated within the linear response regime, with a view to establishing a current-density functional formalism for an inhomogeneous superfluid state. After evaluating in full detail a simplified case of an external coupling to the density and phase of the condensate, the theory is extended to include the coupling to the total current density. The Kohn-Sham response functions of the condensate and all the exchange-correlation kernels for the superfluid are introduced from the microscopic equations of motion and are expressed in a physically transparent way through functional derivatives of correlation functions. A microscopic formula for the superfluid density is derived and used to introduce a generalized hydrodynamic approach for a weakly inhomogeneous two-fluid model in isothermal conditions. Local-density expressions are thereby derived for the velocities of first and second sound in the weakly inhomogeneous superfluid and for visco-elastic functions describing the transition from the hydrodynamic to the collisionless regime. Landau's hydrodynamic theory and known results in Green's functions language are recovered in the limiting case of a homogeneous superfluid.

* Corresponding author. Tel.: +39 50 509058; fax: +39 50 563513; e-mail: minguzzi@cibs.sns.it.

# 1. Introduction

A considerable experimental effort is currently being devoted to the elucidation of the dynamical behaviour of dilute alkali vapours which have undergone Bose-Einstein condensation in magnetic traps. These experimental studies have concerned the excitation of low-lying shape-deformation modes [1 - 3], the propagation of sound waves in the condensate and its thermal cloud [4, 5] and antisymmetric oscillations of the condensate and the thermal cloud corresponding to second sound [5]. The experimental progress has stimulated a number of theoretical approaches involving quantal hydrodynamic descriptions of confined Bose-condensed fluids [6 - 11]. The relationship between the dynamics of such inhomogeneous fluids and the well-known dynamic behaviour of a homogeneous Bose superfluid [12 - 14] also is of considerable interest.

The developments that we have recalled above have motivated the present theoretical study. We bring forward the proposal made in our earlier work [7] to extend to a superfluid of neutral Bose particles the recent progress made in developing a time-dependent current functional theory for the dynamics of inhomogeneous electron systems. For a many-electron system in the normal state the basis of the theory comes from a set of theorems by Runge and Gross [15], showing that the problem of its dynamical behaviour in an external time-dependent potential can be mapped into that of non-interacting electrons in a self-consistent effective potential. These general theorems have been extended to inhomogeneous superconductors [16], the central result being a set of time-dependent Bogolubov - de Gennes equations which include exchange and correlation (xc) effects. In actual applications to electronic systems in the normal state (for reviews see Gross and Kohn[17] and Gross *et al.* [18]) low-frequency phenomena have been usefully described by means of the so-called "adiabatic local density approximation" (ALDA), in which the dynamic xc potential is evaluated as in the corresponding static problem from the xc energy density of the homogeneous electron gas at the local time-dependent density. More recently the search for a fully dynamic local-density approximation has led Vignale and Kohn [19, 20] to develop a current-density formulation of the theory. In brief, these authors have shown that the linear response of a weakly inhomogeneous electron system to an external time-dependent potential can be expressed in terms of (i) a dynamic xc *vector* potential built from xc kernels which are taken from the homogeneous fluid at the local equilibrium density, and (ii) a Kohn-Sham current response matrix to be evaluated from a suitable set of single-particle orbitals. Such a framework supplements the ALDA in embodying not only the electron-gas compressibility but also plasmon dispersion and damping as well as transverse-current fluctuations [21, 22] and allows a unified treatment of the damping of collective excitations from the Landau and mode-coupling mechanisms. The formalism can be readily reduced to yield generalized hydrodynamic equations showing that the xc kernels have the meaning of frequency-dependent viscosity coefficients and elastic moduli [23].

In the case of a normal fluid at zero temperature as treated by Vignale and Kohn [19] the only relevant dynamical variable is the current density. For an extension of this approach to a superfluid of neutral Bose particles in isothermal conditions at finite temperature it is immediately realized that the set of basic dynamical variables must include, in addition to the total current density, the density of the condensate and its phase, the gradient of the latter giving the superfluid velocity field below threshold for vortex generation. In section 2 we present in full detail the theory of the linear response of the two dynamical variables of an inhomogeneous condensate to a dynamic gauge-breaking external field, starting from the microscopic equation of motion for the order parameter in terms of the condensate self-energy. This response is explicitly shown to have the Hohenberg-Kohn-Sham structure compatible with a mapping of the interacting system into a single-particle reference system. We thus identify the Kohn-Sham response functions of the condensate as well as the relevant xc kernels, the latter being expressed in the form of functional derivatives of correlation functions. The connection between the condensate kernels and the Green's functions which are more commonly used to describe the dynamics of homogeneous superfluids [12, 24-27] is exposed in Appendix A.

The response of the noncondensate and of the total current density, and the cross-couplings between condensate and noncondensate are then evaluated in section 3 and in Appendix B. A useful result which emerges from this treatment is a microscopic expression for the superfluid density. This introduces a two-fluid model for the generalized hydrodynamics of an inhomogeneous superfluid in the linear-response, weak inhomogeneity regime, which is developed in § 4. The crucial step is an extension of Landau's linearized hydrodynamic theory for the homogeneous superfluid [13] to finite-frequency phenomena, on the basis of the memory function formalism [28]. The basic assumption of such a theory is that both the equilibrium density profile and the external perturbing fields are slowly varying in space, on length scales set by the interatomic distance and by $c/\omega$ where $c$ is the local speed of sound and $\omega$ is the excitation frequency. Section 4 also presents a brief discussion of the evaluation of the xc kernels for a weakly inhomogeneous superfluid in a collisionless regime. Finally, section 5 gives a summary of our main results.

## 2. Response of the condensate

We treat in this section the linear response of an inhomogeneous condensate to an external perturbation described by the Hamiltonian

$$H'(t) = \int d^3r [\eta(\mathbf{r},t)\psi^\dagger(\mathbf{r}) + \eta^*(\mathbf{r},t)\psi(\mathbf{r})] \qquad (2.1)$$

where $\psi(\mathbf{r})$ is the field operator and $\eta(\mathbf{r},t)$ is a symmetry-breaking scalar field. An account of the response of the noncondensate is deferred to section 3.

*2.1 Coupling to fluctuations in condensate density and superfluid velocity.*

As a first step in making explicit the physical meaning of the Hamiltonian (2.1), we follow Hohenberg and Martin [12] in introducing the transformation

$$\psi^\dagger(\mathbf{r}) = [\hat{n}_c(\mathbf{r})]^{1/2} \exp[-i\hat{\varphi}(\mathbf{r})] \quad, \qquad (2.2)$$

with $\hat{n}_c(\mathbf{r})$ the density operator and $\hat{\varphi}(\mathbf{r})$ the phase operator of the condensate. Then, writing $\psi(\mathbf{r}) = <\psi(\mathbf{r})>_{eq} + \delta\psi(\mathbf{r})$ and with the definition $n_c(\mathbf{r}) = |<\psi(\mathbf{r})>_{eq}|^2$ for the equilibrium density of the condensate, eqn (2.1) can be rewritten as

$$H'(t) = \int d^3r [\alpha(\mathbf{r},t)\delta\hat{n}_c(\mathbf{r}) + \vartheta(\mathbf{r},t)\delta\hat{\varphi}(\mathbf{r})] \qquad (2.3)$$

where $\delta\hat{n}_c(\mathbf{r}) = 2\text{Re}[<\psi(\mathbf{r})>_{eq} \delta\psi^\dagger(\mathbf{r})]$ and $\delta\hat{\varphi}(\mathbf{r}) = -\text{Im}[\delta\psi^\dagger(\mathbf{r})/<\psi^\dagger(\mathbf{r})>_{eq}]$. The fields $\alpha(\mathbf{r},t) = [n_c(\mathbf{r})]^{-1}\text{Re}[<\psi^\dagger(\mathbf{r})>_{eq}\eta(\mathbf{r},t)]$ and $\vartheta(\mathbf{r},t) = 2\text{Im}[<\psi^\dagger(\mathbf{r})>_{eq}\eta(\mathbf{r},t)]$ in eqn (2.3) evidently represent external couplings to fluctuations in the density and phase of the condensate. We use the symbol $<\cdots>$ to denote expectation values on the equilibrium ensemble at given temperature, with the suffix *eq* added to denote properties of the unperturbed fluid.

The superfluid velocity operator $\hat{\mathbf{v}}_s(\mathbf{r})$ is introduced as

$$\hat{\mathbf{v}}_s(\mathbf{r}) = m^{-1}\nabla_\mathbf{r}\delta\hat{\varphi}(\mathbf{r}) \qquad (2.4)$$

with $m$ the particle mass [12]. We are assuming that the superfluid velocity vanishes everywhere at equilibrium. The perturbing Hamiltonian (2.3) is then rewritten as

$$H'(t) = \int d^3r [\alpha(\mathbf{r},t)\delta\hat{n}_c(\mathbf{r}) + \boldsymbol{\lambda}(\mathbf{r},t)\cdot\hat{\mathbf{v}}_s(\mathbf{r})] \quad , \tag{2.5}$$

where

$$\boldsymbol{\lambda}(\mathbf{r},t) = \frac{m}{4\pi}\int d^3r' \frac{1}{|\mathbf{r}-\mathbf{r}'|}\nabla_{\mathbf{r}'}\vartheta(\mathbf{r}',t) \quad . \tag{2.6}$$

In summary, the Hamiltonian (2.1) describes external fields acting on the density and phase fluctuations of an inhomogeneous condensate, the gradient of the phase fluctuations being related by eqn (2.4) to fluctuations in the superfluid velocity.

*2.2 Linear response functions and their connection to Green's functions.*

A two-by-two matrix of (time-ordered) response functions for the condensate density and phase is introduced through the definitions

$$\chi_{\varphi\varphi}(1,1') \equiv \frac{\delta<\hat{\varphi}(1)>}{\delta\vartheta(1')} = <T[\delta\hat{\varphi}(1)\delta\hat{\varphi}(1')]> \tag{2.7}$$

etcetera. Here, $1 \equiv (\mathbf{r}_1, t_1)$ and the operators are in the Heisenberg representation.

With the notation $\psi(1) = \psi_1(1)$ and $\psi^\dagger(1) = \psi_2(1)$, we introduce symmetrized and equilibrium-weighed single-particle Green's functions $\overline{G}^{(\pm)}_{\alpha\beta}(1,1')$ for the noncondensate as

$$\overline{G}^{(\pm)}_{\alpha\beta}(1,1') = \frac{1}{2}[-<\psi^\dagger_\alpha(\mathbf{r}_1)>_{eq} \tilde{G}_{\alpha\beta}(1,1') <\psi_\beta(\mathbf{r}_{1'})>_{eq} \text{m}c.c.] \tag{2.8}$$

where

$$\tilde{G}_{\alpha\beta}(1,1') = -i\{<T[\psi_\alpha(1)\psi^\dagger_\beta(1')]> - <\psi_\alpha(1)><\psi^\dagger_\beta(1')>\} \quad . \tag{2.9}$$

The following expressions are then obtained for the time-ordered response functions from the definitions of $\delta\hat{n}_c(\mathbf{r})$ and $\delta\hat{\varphi}(\mathbf{r})$:

$$\chi_{\varphi\varphi}(1,1') = i[2n_c(\mathbf{r}_1)n_c(\mathbf{r}_{1'})]^{-1}[\overline{G}^{(+)}_{11}(1,1') - \overline{G}^{(+)}_{12}(1,1')] \quad , \tag{2.10}$$

$$\chi_{n_c n_c}(1,1') = 2i[\overline{G}^{(+)}_{11}(1,1') + \overline{G}^{(+)}_{12}(1,1')] \quad , \tag{2.11}$$

$$\chi_{n_c \varphi}(1,1') = -[n_c(\mathbf{r}_{1'})]^{-1}[\overline{G}^{(-)}_{11}(1,1') - \overline{G}^{(-)}_{12}(1,1')] \tag{2.12}$$

and

$$\chi_{\varphi n_c}(1,1') = [n_c(\mathbf{r}_1)]^{-1}[\overline{G}^{(-)}_{11}(1,1') + \overline{G}^{(-)}_{12}(1,1')] \quad . \tag{2.13}$$

Of course, the expressions (2.10), (2.12) and (2.13) can be converted with the help of eqn (2.4) into expressions for response functions involving the superfluid velocity.

In the present case the off-diagonal response functions coupling the superfluid velocity to the condensate density reflect the non-conservation of the number of particles in the condensate: by exciting fluctuations in the condensate one changes its density and hence induces a superfluid flow. Whereas in standard treatments of superfluid hydrodynamics the amplitude fluctuations of the condensate can be neglected (see e. g. [14]), all the four response functions (2.10) - (2.13) are taken into account in our treatment, which is not restricted to low frequency.

The symmetry properties of the response functions introduced above are easily assessed by a standard analysis of the spectral functions associated with the causal analogues of the Green's functions (2.8). Assuming a time-inversion-invariant unperturbed state, one can in particular prove the symmetry relation $\text{Im}\,\chi_{\varphi n_c}(\mathbf{r},\mathbf{r}';\omega) = -\text{Im}\,\chi_{n_c\varphi}(\mathbf{r}',\mathbf{r};-\omega)$, both functions being purely imaginary and even in $\omega$.

*2.3 Microscopic expressions for the response functions.*

We proceed to evaluate the response functions from the microscopic equation of motion for the expectation value $<\psi(1)>$ of the field operator in the presence of the perturbation $H'(t)$. This is

$$\int d\bar{2}\, G_0^{-1}(1,\bar{2}) <\psi(\bar{2})> = \eta(1) + \sigma(1) \tag{2.14}$$

where $G_0$ is the free-particle Green's function,

$$G_0^{-1}(1,1') = [i\frac{\partial}{\partial t_1} + \frac{1}{2m}\nabla_{\mathbf{r}_1}^2 - V(\mathbf{r}_1) + \mu]\delta(1,1') \tag{2.15}$$

with $V(\mathbf{r}_1)$ the confining potential and $\mu$ the chemical potential, and

$$\sigma(1) = \int d\bar{2}\, v(\mathbf{r}_1,\mathbf{r}_{\bar{2}}) <\psi^\dagger(\bar{2})\psi(\bar{2})\psi(1)> \tag{2.16}$$

is the condensate self-energy, with $v(\mathbf{r}_1,\mathbf{r}_{1'})$ the interparticle pair potential.

By analogy with the definitions of the external fields entering eqn (2.3), we introduce the excess (ex) potentials

$$\alpha_{ex}(1) = [n_c(\mathbf{r}_1)]^{-1}\text{Re}[<\psi^\dagger(\mathbf{r}_1)>_{eq}\sigma(1)] \tag{2.17}$$

and

$$\vartheta_{ex}(1) = 2\,\text{Im}[<\psi^\dagger(\mathbf{r}_1)>_{eq}\sigma(1)]\quad. \tag{2.18}$$

The ex kernel $\alpha_{ex}$ may be viewed as the shift in chemical potential due to the interactions in the fluid away from equilibrium. On the other hand, $\vartheta_{ex}$ determines the violation of the continuity equation for the condensate through couplings to the noncondensate, as may be seen from the microscopic equation of motion for the condensate density fluctuations (see also Hohenberg and Martin [12]).

Equation (2.14) can then be written as

$$<\psi(1)> = \int d\bar{2}\, G_0(1,\bar{2}) <\psi^\dagger(\mathbf{r}_{\bar{2}})>_{eq}\{[\alpha(\bar{2}) + \alpha_{ex}(\bar{2})]$$

$$+\frac{1}{2}i[n_c(\mathbf{r}_{\bar{2}})]^{-1}[\vartheta(\bar{2})+\vartheta_{ex}(\bar{2})]\} \quad . \quad (2.19)$$

This form of the equation of motion for the field operator can now be used to evaluate the response functions by taking functional derivatives with respect to the external fields in the linear response limit.

Let us consider first the phase-phase response function $\chi_{\varphi\varphi}(1,1')$. From its definition we have

$$\chi_{\varphi\varphi}(1,1') = [in_c(\mathbf{r}_1)]^{-1}\int d\bar{2}\{\frac{1}{2}i\overline{G}_0^{(+)}(1,\bar{2})[n_c(\mathbf{r}_{\bar{2}})]^{-1}[\delta(\bar{2},1')+\frac{\delta\vartheta_{ex}(\bar{2})}{\delta\vartheta(1')}]$$
$$+\overline{G}_0^{(-)}(1,\bar{2})\frac{\delta\alpha_{ex}(\bar{2})}{\delta\vartheta(1')}\} \quad (2.20)$$

where

$$\overline{G}_0^{(\pm)}(1,1') = \frac{1}{2}[<\psi^\dagger(\mathbf{r}_1)>_{eq} G_0(1,1') <\psi(\mathbf{r}_{1'})>_{eq} \pm c.c.] \quad . \quad (2.21)$$

It is evident from eqn (2.20) that $\overline{G}_0^{(+)}$ is related to an ideal phase-phase response function $\chi_{\varphi\varphi}^{(0)}$, aside from weighing factors determined by the equilibrium density of the real fluid:

$$\chi_{\varphi\varphi}^{(0)}(1,1') = i[2n_c(\mathbf{r}_1)n_c(\mathbf{r}_{1'})]^{-1}\overline{G}_0^{(+)}(1,1') \quad . \quad (2.22)$$

A similar calculation for the phase-density response function attributes to $\overline{G}_0^{(-)}$ in eqn (2.21) the meaning of an ideal phase-density response function,

$$\chi_{\varphi n_c}^{(0)}(1,1') = [n_c(\mathbf{r}_1)]^{-1}\overline{G}_0^{(-)}(1,1') \quad . \quad (2.23)$$

Equation (2.20) can therefore be written in its final form,

$$\chi_{\varphi\varphi} = \chi_{\varphi\varphi}^{(0)} + [\chi_{\varphi\varphi}^{(0)} \otimes \frac{\delta\vartheta_{ex}}{\delta\varphi}|_{n_c} + \chi_{\varphi n_c}^{(0)} \otimes \frac{\delta\alpha_{ex}}{\delta\varphi}|_{n_c}] \otimes \chi_{\varphi\varphi}$$
$$+[\chi_{\varphi\varphi}^{(0)} \otimes \frac{\delta\vartheta_{ex}}{\delta n_c}|_\varphi + \chi_{\varphi n_c}^{(0)} \otimes \frac{\delta\alpha_{ex}}{\delta n_c}|_\varphi] \otimes \chi_{n_c\varphi} \quad . \quad (2.24)$$

In eqn (2.24) we have omitted the arguments of all the functions involved and indicated with the symbol $\otimes$ an integration over intermediate variables. Of course, eqn (2.4) immediately yields

$$\chi_{\mathbf{v}_s\mathbf{v}_s}(1,1') = m^{-2}\nabla_{\mathbf{r}_1}\nabla_{\mathbf{r}_{1'}}\chi_{\varphi\varphi}(1,1') \quad . \quad (2.25)$$

The microscopic structure of the matrix of response functions is now clear. With the further definitions

$$\chi_{n_c\varphi}^{(0)}(1,1') = -[n_c(\mathbf{r}_1)]^{-1}\overline{G}_0^{(-)}(1,1') \quad (2.26)$$

and

$$\chi_{n_c n_c}^{(0)}(1,1') = 2i\overline{G}_0^{(+)}(1,1') \quad (2.27)$$

we construct an ideal response matrix $\chi^{(0)}$:

$$\chi^{(0)} = \begin{pmatrix} \chi^{(0)}_{n_c n_c} & \chi^{(0)}_{n_c \varphi} \\ \chi^{(0)}_{\varphi n_c} & \chi^{(0)}_{\varphi \varphi} \end{pmatrix} \quad . \tag{2.28}$$

We also have the matrix $K$ of excess kernels arising from the interactions:

$$K = \begin{pmatrix} \delta \alpha_{ex} / \delta n_c |_\varphi & \delta \alpha_{ex} / \delta \varphi |_{n_c} \\ \delta \vartheta_{ex} / \delta n_c |_\varphi & \delta \vartheta_{ex} / \delta \varphi |_{n_c} \end{pmatrix} \quad . \tag{2.29}$$

Then the desired form of the matrix $\chi$ of response functions in the real Bose fluid is given by

$$\chi = \chi^{(0)} + \chi^{(0)} \otimes K \otimes \chi \tag{2.30}$$

or

$$\chi = [I - \chi^{(0)} \otimes K]^{-1} \otimes \chi^{(0)} \quad . \tag{2.31}$$

As a final remark we explicitly note that, at variance from what is customary for electron fluids, the kernels $\delta \alpha_{ex} / \delta n_c |_\varphi$ and $\delta \alpha_{ex} / \delta \varphi |_{n_c}$ in eqn (2.29) include Hartree contributions in addition to the exchange and correlation terms.

The relevance of equation (2.30) to a density-functional treatment of the dynamics of an inhomogeneous superfluid will be discussed in section 2.4 immediately below. An alternative derivation of eqn (2.30), which highlights the meaning of the excess kernels (2.29) as single-particle self-energies for the real fluid and relates the present treatment to the work of Wong and Gould [25] on homogeneous Bose fluids, is given in Appendix A.

*2.4 Connection with time-dependent density functional theory*

The structure of the microscopic equations (2.30) for the response of the condensate at fixed noncondensate is a consequence of the fact that the self-energy $\sigma(1)$ is a function of only one space-time variable. In our derivation of these equations in § 2.3 we chose to use the response functions (2.28) as built from the ideal-gas Green's function $G_0$, in order to facilitate comparisons with earlier work on the homogeneous fluid. Hence, the form of eqn (2.30) is that appropriate for a Hohenberg-Kohn-Sham treatment, with a Kohn-Sham reference fluid which is, however, an ideal Bose gas except for weighing factors determined by the true equilibrium condensate density.

A more appropriate set of Kohn-Sham response functions for a time-dependent density functional treatment of the condensate response can be obtained from a single-particle Green's function $G_{KS}(1,1')$ defined through

$$G_{KS}^{-1}(1,1') = G_0^{-1}(1,1') - \frac{\sigma_{eq}(\mathbf{r}_1)}{<\psi(\mathbf{r}_1)>_{eq}} \delta(1,1') \quad . \tag{2.32}$$

The equilibrium self-energy term which has been included in eqn (2.32) needs to be subtracted from the RHS of eqn (2.14), which now reads

$$\int d\bar{2} G_{KS}^{-1}(1,\bar{2}) <\psi(\bar{2})> = \eta(1) + \Delta\sigma(1) \qquad (2.33)$$

where

$$\Delta\sigma(1) = \sigma(1) - \sigma_{eq}(\mathbf{r}_1) <\psi(1)> / <\psi(\mathbf{r}_1)>_{eq} \qquad (2.34)$$

contains only dynamical effects. All other equations of § 2.3 remain formally the same, after replacement of $G_0$ with $G_{KS}$ and of $\sigma(1)$ with $\Delta\sigma(1)$. In particular, the matrix of Kohn-Sham response functions $\chi_{KS}$ is to be built from the Green's functions

$$\overline{G}_{KS}^{(\pm)}(1,1') = \frac{1}{2}[<\psi^\dagger(\mathbf{r}_1)>_{eq} G_{KS}(1,1') <\psi(\mathbf{r}_{1'})>_{eq} \pm c.c.] \qquad (2.35)$$

and is in general non-diagonal. We also remark that a static mean-field potential due to the interactions has been included in earlier treatments of the dynamics of an inhomogeneous condensate in the random phase approximation [29, 30].

Thus, the Kohn-Sham reference system that we are proposing for a density functional approach to the dynamical xc effects in a superfluid explicitly contains the interactions through the equilibrium value of the condensate self-energy. Of course, expansion of the Kohn-Sham response functions in an appropriate basis set will be needed (see *e.g.* the suggestion of a Bogolubov - de Gennes basis set made by Wacker *et al.* [16] for a superconductor). On the other hand, we should emphasize that we are not advocating a density functional approach to the evaluation of the equilibrium state of the superfluid. In particular, a thermodynamic treatment based on the Gross-Pitaevskii equation is in quantitative agreement with experiment for confined vapours of alkali atoms which have undergone Bose-Einstein condensation [31].

We conclude this section by recalling that simple approximations on $\Delta\sigma(1)$, leading to a time-dependent Gross-Pitaevskii equation, have been found to be useful in the recent literature on collective excitations of the condensate (see *e.g.* [32]).

## 3. Response of the noncondensate and of the whole fluid

Having dealt with the linear response properties of the condensate in the preceding section, we now extend the treatment to the linear response of the whole Bose fluid. The perturbing Hamiltonian (2.1) is supplemented by the perturbation

$$H''(t) = \int d^3r \int d^3r' \psi^\dagger(\mathbf{r}')U(\mathbf{r},\mathbf{r}';t)\psi(\mathbf{r}) \quad , \qquad (3.1)$$

involving a non-local scalar potential $U(\mathbf{r},\mathbf{r'};t)$ as needed to describe the current response. We shall first focus on the response of the noncondensate to this potential, before turning to treat the response of the whole fluid to the set of external perturbations given by $\eta(\mathbf{r},t)$ and $U(\mathbf{r},\mathbf{r'};t)$.

In earlier work on Bose fluids by the so-called dielectric formalism [12, 14, 24, 25], a diagrammatic analysis has often been used to divide the proper current response function into a 'regular' part and a 'singular' part. The regular part is described by the set of irreducible diagrams and may be identified with the response of the normal component. The singular part can be expressed in terms of the response of the condensate, multiplied by a vertex function which contains off-diagonal (condensate-noncondensate) kernels. The same vertex function relates the condensate-noncondensate response to that of the condensate.

Here, instead, we evaluate the total current response as the sum of the response of the noncondensate at fixed condensate and of the response of the condensate to the $U$ field. Both contributions are obtained by a functional-derivative technique from the equations of motion of the single-particle Green's functions $\tilde{G}$. In our calculations we take the single-particle self-energies as functionals of $\tilde{G}$ and of the order parameter $<\psi>$, this approach being the most general and naturally amenable to perturbative treatments [33]. The results can be immediately compared with those of Griffin [14], allowing us to identify the two contributions to the total current response as coming from the irreducible and from the reducible diagrams, respectively.

*3.1 Linear response of the noncondensate.*

We define the response function for the current density of the noncondensate through the functional derivative of its Green's function with respect to the potential $U$ at fixed condensate (*i.e.* with the changes in condensate density and superfluid velocity set equal to zero):

$$\chi_{\tilde{\mathbf{j}}\tilde{\mathbf{j}}}(1,2) = i(2m)^{-2}[\nabla_{1,1'}\nabla_{2,2'}\frac{\delta\tilde{G}_{11}(1,1')}{\delta U(2,2')}|_{(n_c,v_s)}]_{1=1',2=2'} \quad , \qquad (3.2)$$

where $\nabla_{1,1'} = \nabla_1 - \nabla_{1'}$. The equation for the general four-point response function $\delta\tilde{G}_{\alpha\beta}/\delta U$ is obtained by taking a functional derivative on the Dyson equation for $\tilde{G}_{\alpha\beta}$:

$$\frac{\delta\tilde{G}_{\alpha\beta}(1,1')}{\delta U(2,2')} = \tilde{G}_{\alpha\gamma}(1,2)\tilde{G}_{\gamma\beta}(2',1') + \tilde{G}_{\alpha\gamma}(1,\bar{3})\frac{\delta\Sigma_{\gamma\delta}(\bar{3},\bar{3}')}{\delta U(2,2')}\tilde{G}_{\delta\beta}(\bar{3}',1') \quad , \qquad (3.3)$$

where $\Sigma_{\alpha\beta}$ are the single-particle self-energies,

$$\Sigma_{\alpha\beta}(1,1') = \frac{\delta\sigma_\alpha(1)}{\delta<\psi_\beta(1')>}|_U \quad , \tag{3.4}$$

with $\sigma_1(1) = \sigma(1)$ and $\sigma_2(1) = \sigma^*(1)$ [34]. The convention of integration over repeated barred variables and of summation over repeated Greek indices has been used. Equations (3.3) and (3.4) are used in Appendix B to derive a microscopic expression for $\chi_{\tilde{j}\tilde{j}}$ (see eqns (B.1) - (B.3)).

Of special interest here is the relationship between $\chi_{\tilde{j}\tilde{j}}$ and the noncondensate density-current response $\chi_{\tilde{n}\tilde{j}}$. From the Dyson equation for $\tilde{G}$ we find the equation of motion for the noncondensate density $\tilde{n}(1) = i\tilde{G}_{11}(1,1^+)$, and hence by taking its functional derivative with respect to $U$ at fixed condensate we get the Ward identity

$$\chi_{\tilde{n}\tilde{j}}(1,1') = (i\omega)^{-1}[\nabla_1 \cdot \chi_{\tilde{j}\tilde{j}}(1,1') + m^{-1}\nabla_1\rho_n(1,1')] \quad . \tag{3.5}$$

In eqn (3.5) we have defined

$$\nabla_1\rho_n(1,2) = \nabla_1[\tilde{n}(1)\delta(1,2)] - \text{Re}\{\nabla_{2,2'}[<\psi^\dagger(1)>_{eq}\frac{\delta\sigma(1)}{\delta U(2,2')}]_{2=2'}\} \quad . \tag{3.6}$$

With the further definition

$$\rho_s(1,2) \equiv n(1)\delta(1,2) - \rho_n(1,2) \quad , \tag{3.7}$$

eqn (3.6) can be rewritten in terms of the vertex functions introduced in Appendix B as

$$\nabla_1\rho_s(1,2) = -m\overset{\perp}{\vartheta}_\Lambda(1,2) \tag{3.8}$$

(see eqn (B.9)). Equations (3.6) and (3.8) extend to the inhomogeneous Bose fluid an equation already derived in the homogeneous case by Huang and Klein [20]. It is easily shown that the function $\rho_s(1,2)$ reduces for the homogeneous fluid in the static limit to the microscopic definition of the superfluid density given by Griffin [14]. Evidently, the function $\rho_n(1,2)$ reduces to the normal-fluid density. We shall make use of these functions in dealing with a two-fluid model in § 4 below.

As a final remark, we notice from eqn (3.5) that in the limit indicated above $\rho_n$ may be interpreted as the so-called diamagnetic contribution to the irreducible current-current response function $\chi_{\tilde{j}\tilde{j}}$ (see also the discussion given by Hohenberg and Martin [12] for the homogeneous Bose fluid).

*3.2 Off-diagonal response and total current-current response.*

In section 3.1 we have treated only the irreducible part of the current-current response. Detailed calculations of the off-diagonal (condensate-noncondensate) response and of the reducible contribution to the current-current response are given in Appendix B. The final result for the current-current response function $\chi_{jj}$ can be cast in a form which is equivalent to eqn (6.38) in the work of Hohenberg and Martin [12]:

$$\chi_{jj} = \chi_{\tilde{j}\tilde{j}} + \begin{pmatrix} \dfrac{\delta \mathbf{j}}{\delta v_s} & \dfrac{\delta \mathbf{j}}{\delta n_c} \end{pmatrix} \otimes \begin{pmatrix} \chi_{v_s v_s} & \chi_{v_s n_c} \\ \chi_{n_c v_s} & \chi_{n_c n_c} \end{pmatrix} \otimes \begin{pmatrix} \dfrac{\delta \mathbf{j}}{\delta v_s} \\ \dfrac{\delta \mathbf{j}}{\delta n_c} \end{pmatrix} \quad . \tag{3.9}$$

A comparison of eqn (3.8) with eqn (B.15), *i.e.*

$$\frac{\delta \mathbf{j}(1)}{\delta \varphi(2)}\bigg|_U = \vec{\vartheta}_\Lambda(1,2) \tag{3.10}$$

yields the result

$$\nabla_1 \rho_s(1,2) = \nabla_2 \cdot \frac{\delta \mathbf{j}(1)}{\delta \mathbf{v}_s(2)}\bigg|_U \quad . \tag{3.11}$$

The microscopic expressions (3.9) and (3.11) should be contrasted with the results reported by Hohenberg and Martin [12] for a two-fluid model subject to slowly varying perturbations, upon neglect of fluctuations in the condensate density (see their eqn (5.16)). These authors show that in such a limit $\chi_{jj}$ is the response of the current density $\mathbf{j}(\mathbf{r},t)$ to an external field given by the normal-fluid velocity $\mathbf{v}_n(\mathbf{r},t)$ and $\chi_{\mathbf{v}_s \mathbf{v}_s}$ is the response of the superfluid velocity $\mathbf{v}_s(\mathbf{r},t)$ to an external field given by the interdiffusion current $\mathbf{j}_r(\mathbf{r},t) \equiv \mathbf{j}(\mathbf{r},t) - n(\mathbf{r})\mathbf{v}_n(\mathbf{r},t) = \rho_s(\mathbf{r})[\mathbf{v}_s(\mathbf{r},t) - \mathbf{v}_n(\mathbf{r},t)]$, $\rho_s(\mathbf{r})$ and $\rho_n(\mathbf{r}) = n(\mathbf{r}) - \rho_s(\mathbf{r})$ being the equilibrium densities of the superfluid and normal-fluid components. Furthermore, their definition of superfluid density involves, as in eqn (3.11), the functional derivative of the current with respect to the superfluid velocity at fixed normal-fluid velocity.

In summary, the main results of this section are the microscopic relations for the generalized normal-fluid and superfluid densities given in eqns (3.6) and (3.11), and the expression (3.9) for the total current-current response. These are supplemented by eqn (B.1) for the irreducible current-current response and eqn (B.7) for the off-diagonal condensate-irreducible current response. We have also made full contact with the two-fluid model as discussed by Hohenberg and Martin [12], for what concerns in particular the superfluid density and the normal-fluid velocity. With regard to the latter we also recall that Kane and Kadanoff [35] have interpreted $\mathbf{v}_n(\mathbf{r},t)$ as the local velocity of bodily flow of the fluid within a microscopic Boltzmann equation approach to the two-fluid hydrodynamic behaviour.

## 4. Current functional formalism for the two-fluid model of a weakly inhomogeneous superfluid

In contrast to the Hohenberg-Kohn-Sham structure that we have derived in § 2 for the linear response of an inhomogeneous Bose condensate, the structure of the current response and cross response functions given in § 3 is quite complex. However, in the last resort one may obtain a practicable calculational scheme from the general formalism only by (i) taking special limiting cases and (ii) introducing approximations. An important limiting case is that of weak inhomogeneity, in which both the unperturbed density profile of the fluid and the external fields acting on it are slowly varying in space. Recent progress for the current response of electronic systems in the normal state [19 - 23] has been referred to in § 1.

Equations (3.5) - (3.7), introducing the functions $\rho_n(1,2)$ and $\rho_s(1,2)$ which for the homogeneous fluid in the static limit reduce to the densities of normal fluid and of superfluid, can be taken as the basis for developing a two-fluid model for the generalized hydrodynamics of the inhomogeneous superfluid in the linear-response, weak-inhomogeneity regime. Specifically, we propose that eqns (3.6) and (3.7) may provide a reasonable definition of the equilibrium superfluid density $\rho_s(\mathbf{r})$ in the weakly inhomogeneous case, when we take its static limit and the $\mathbf{k}=0$ component of its Fourier transform with respect to $\mathbf{r}_1 - \mathbf{r}_2$. The functional derivative entering the RHS of eqn (3.6) is a five-point correlation function, to be evaluated on the equilibrium state.

*4.1 Generalized hydrodynamics of a homogeneous superfluid.*

With this definition of the superfluid density, we first proceed to extend Landau's hydrodynamic theory for the homogeneous superfluid [13] to finite-frequency phenomena. We use for this purpose the well-known memory function formalism, as described *e.g.* in the book of Forster [28]. We assume isothermal conditions, *i.e.* neglect the couplings between density and temperature fluctuations. The form of the generalized hydrodynamic equations is in fact dictated by some general considerations: (i) invariance under a Galileian transformation and Onsager symmetry must hold; (ii) as a consequence of the zero-force and zero-torque theorems, the time derivative of the current density $\mathbf{j}$ is driven by the divergence of a symmetric tensor of the second rank; (iii) the time derivative of the superfluid velocity $\mathbf{v}_s$ is the gradient of a scalar quantity, in view of its irrotational character below threshold for vortex generation; and (iv) as already recalled at the end of § 3, the internal driving forces are determined by the divergence of the normal-fluid velocity $\mathbf{v}_n$ and of the interdiffusion current $\mathbf{j}_r \equiv \mathbf{j} - n\mathbf{v}_n = \rho_s(\mathbf{v}_s - \mathbf{v}_n)$. The form that we propose for the generalized hydrodynamic equations thus is

$$-im\omega\delta\mathbf{j}(\mathbf{r},\omega) = -\vec{\nabla}\cdot[\delta p(\mathbf{r},\omega)\mathbf{\bar{I}}] + \vec{\nabla}\cdot\bar{\sigma}(\mathbf{r},\omega) \qquad (4.1)$$

and

$$-im\omega\delta\mathbf{v}_s(\mathbf{r},\omega) = -\vec{\nabla}[\delta\mu(\mathbf{r},\omega)] + \vec{\nabla}\cdot\bar{\sigma}^{(s)}(\mathbf{r},\omega) \quad , \qquad (4.2)$$

where the stress tensors are given by

$$\sigma_{ij} = [\eta(\omega) - p_0(n)/i\omega](\frac{\partial v_{ni}}{\partial r_j} + \frac{\partial v_{nj}}{\partial r_i} - \frac{2}{3}\delta_{ij}\vec{\nabla}\cdot\mathbf{v}_n)$$
$$+ \delta_{ij}[\zeta_2(\omega)\vec{\nabla}\cdot\mathbf{v}_n + \zeta_1(\omega)\vec{\nabla}\cdot\mathbf{j}_r] \qquad (4.3)$$

and

$$\sigma_{ij}^{(s)} = \delta_{ij}[\zeta_3(\omega)\vec{\nabla}\cdot\mathbf{j}_r + \zeta_4(\omega)\vec{\nabla}\cdot\mathbf{v}_n] \quad . \qquad (4.4)$$

Here, $p_0(n)$ is the ideal-gas pressure at the equilibrium density $n$, while $\delta p(\mathbf{r},\omega)$ and $\delta\mu(\mathbf{r},\omega)$ are the local pressure and chemical potential fluctuations. These are given in terms of the density fluctuations $\delta n(\mathbf{r},\omega)$ and of the entropy fluctuations $\delta s(\mathbf{r},\omega)$ by the linearized expressions

$$\delta p(\mathbf{r},\omega) = (nK_T)^{-1}\delta n(\mathbf{r},\omega) \qquad (4.5)$$

and

$$\delta\mu(\mathbf{r},\omega) = (n^2 K_T)^{-1}\delta n(\mathbf{r},\omega) + (Ts/c_V)\delta s(\mathbf{r},\omega) \quad . \qquad (4.6)$$

Here, $K_T$, $s$ and $c_V$ are the isothermal compressibility, the entropy and the constant-volume specific heat (per particle). Of course, in isothermal conditions the entropy fluctuations are associated with fluctuations in the density of the noncondensate.

The functions $\zeta_i$ and $\eta$ in eqns (4.1) - (4.4) are frequency-dependent visco-elastic coefficients, whose real parts express attenuation spectra. Within the memory function formalism these spectra are related to the response functions of the system by the following generalized Kubo formulae:

$$\mathrm{Re}[\zeta_2(\omega) + \frac{4}{3}\eta(\omega)] = \lim_{k \to 0} \frac{-\omega m^2}{k^2} \mathrm{Im}\,\chi^L_{jj}(k,\omega) \quad , \tag{4.7}$$

$$\mathrm{Re}\,\eta(\omega) = \lim_{k \to 0} \frac{-\omega m^2}{k^2} \mathrm{Im}\,\chi^T_{jj}(k,\omega) \quad , \tag{4.8}$$

$$\mathrm{Re}\,\zeta_3(\omega) = \lim_{k \to 0} \frac{-\omega}{k^2} \mathrm{Im}\,\chi_{v_s v_s}(k,\omega) \tag{4.9}$$

and

$$\mathrm{Re}[\zeta_1(\omega)] = \mathrm{Re}[\zeta_4(\omega)] = \lim_{k \to 0} \frac{-\omega m}{k^2} \mathrm{Im}\,\chi^L_{jv_s}(k,\omega) \quad , \tag{4.10}$$

the equality $\zeta_1 = \zeta_4$ being due to Onsager symmetry. Of course, the imaginary parts of these spectra, which have the meaning of finite-frequency elastic moduli, are related to the real parts by Kramers-Kronig relations.

The next step involves relating the visco-elastic functions to ex kernels introduced from the long-wavelength behaviour of the $\mathbf{j}$ - $\mathbf{v}_s$ response functions and of their single-particle equivalents. Within the two-fluid model, *i.e.* setting $\mathbf{j} = \rho_s \mathbf{v}_s + \rho_n \mathbf{v}_n$ and $n = \rho_s + \rho_n$, we use the results reported in eqns (4.24) - (4.26) in the work of Hohenberg and Martin [12] and those given in Table 2.1 of the book of Nozières and Pines [36] to find the following limiting behaviours:

$$\lim_{k \to 0} \chi_{v_\alpha v_\beta}(k,\omega) = (\rho_\alpha)^{-1}\delta_{\alpha\beta} + A_{\alpha\beta}(\omega)k^2 + o(k^2) \tag{4.11}$$

and

$$\lim_{k \to 0} \chi^0_{v_\alpha v_\beta}(k,\omega) = (\rho_\alpha)^{-1}\delta_{\alpha\beta} + A^0_{\alpha\beta}(\omega)k^2 + o(k^2) \quad . \tag{4.12}$$

Here, the Greek indices refer to the superfluid (s) and to the normal (n) component, $\chi^0_{v_\alpha v_\beta}$ are the single-particle response functions and the $A$´s are frequency-dependent coefficients. The ex kernels are then defined through the equation

$$\overset{t}{\mathbf{f}}_{\alpha\beta}(\omega) = \lim_{k \to 0} \frac{\omega^2}{k^2} \rho_\alpha \rho_\beta [\chi_{v_\alpha v_\beta}(k,\omega) - \chi^0_{v_\alpha v_\beta}(k,\omega)] \quad . \tag{4.13}$$

Use of the relations (4.7) - (4.10) and of their Hilbert transforms yields the following expressions for the visco-elastic coefficients in terms of the ex kernels (4.13):

$$\zeta_1(\omega;n,T) = -(i\omega)^{-1}[f_{j_L v_s}(\omega;\{\rho_\alpha\}) - \partial p_{ex}(n,T)/\partial n|_T] \quad , \tag{4.14}$$

$$\zeta_2(\omega;n,T) = -(i\omega)^{-1}[f_{j_L j_L}(\omega;\{\rho_\alpha\}) - 4f_{j_T j_T}(\omega;\{\rho_\alpha\})/3$$
$$- n\partial p_{ex}(n,T)/\partial n|_T] \quad , \tag{4.15}$$

$$\zeta_3(\omega;n,T) = -(i\omega)^{-1}[f_{v_s v_s}(\omega;\{\rho_\alpha\}) - \partial \mu_{ex}(n,T)/\partial n|_T$$
$$-(Ts/c_V)\partial \mu_{ex}(n,T)/\partial T|_n] \quad , \tag{4.16}$$

$$\zeta_4(\omega;n,T) = -(i\omega)^{-1}[f_{v_s j_L}(\omega;\{\rho_\alpha\}) - \partial p_{ex}(n,T)/\partial n|_T] \tag{4.17}$$

and

$$\eta(\omega;n,T) = -(i\omega)^{-1} f_{j_T j_T}(\omega;\{\rho_\alpha\}) \quad . \tag{4.18}$$

Here, $p_{ex}$ and $\mu_{ex}$ are the excess pressure and chemical potential. We have denoted by $\{\rho_\alpha\} = (\rho_s, \rho_n)$ the densities in the equilibrium state and left implicit the temperature dependence of the ex kernels.

*4.2. Local-density theory of the weakly inhomogeneous superfluid.*

The above results are easily extended to a weakly inhomogeneous superfluid in isothermal conditions. The general properties that we have used above to derive the structure of the generalized hydrodynamic equations in the homogeneous case remain valid, while a Ward identity is essential to relate the effect of a weak inhomogeneity on the excess kernels to their density dependence. This identity is obtained by modulating the densities and comparing the long-wavelength behaviour of the inhomogeneous response functions, to first order in the inhomogeneity, with those of the modulated homogeneous system [19].

Following the steps given in the derivation of Vignale and Kohn [20] for the electron fluid, we create a weak modulation of the densities of the superfluid and normal-fluid components given by $\delta\rho_\alpha(\mathbf{r}) = 2\xi_\alpha \bar{\rho}_\alpha \cos(\mathbf{q}\cdot\mathbf{r})$ at long wavelengths $\mathbf{q}\to 0$, with $\xi_\alpha \ll 1$. The ex kernels in such an inhomogeneous system can be written in the form

$$\breve{\mathbf{f}}^{inh}_{\alpha\beta}(\mathbf{k},\mathbf{r};\omega) = \breve{\mathbf{f}}^{h}_{\alpha\beta}(\mathbf{k},\omega;\{\bar{\rho}_\alpha\}) + 2\breve{\mathbf{f}}_{\alpha\beta}(\mathbf{k}+\mathbf{q},\mathbf{k},\omega;\{\bar{\rho}_\alpha\})\cos(\mathbf{q}\cdot\mathbf{r}) \tag{4.19}$$

at long wavelengths $\mathbf{k}\to 0$. The Ward identity reads

$$\lim_{\mathbf{q}\to 0} \overset{\leftrightarrow}{\mathbf{f}}_{\alpha\beta}(\mathbf{k}+\mathbf{q},\mathbf{k},\omega;\{\bar{\rho}_\alpha\}) = \sum_\gamma \xi_\gamma \bar{\rho}_\gamma \frac{\partial}{\partial \bar{\rho}_\gamma} \overset{\leftrightarrow}{\mathbf{f}}^h_{\alpha\beta}(\mathbf{k},\omega;\{\bar{\rho}_\alpha\}) \qquad (4.20)$$

and hence eqn (4.19) yields

$$\overset{\leftrightarrow}{\mathbf{f}}^{inh}_{\alpha\beta}(\mathbf{k},\mathbf{r},\omega) = \overset{\leftrightarrow}{\mathbf{f}}^h_{\alpha\beta}(\mathbf{k},\omega;\{\rho_\alpha(\mathbf{r})\}) \quad . \qquad (4.21)$$

Namely, the kernels in the weakly inhomogeneous superfluid are given by those of the homogeneous superfluid taken at the local equilibrium densities.

In summary, the combination of the zero-force and zero-torque theorems with the Ward identity (4.21) determines the generalized hydrodynamic equations of the weakly inhomogeneous superfluid in the form

$$-im\omega\delta\mathbf{j}(\mathbf{r},\omega) = -n(\mathbf{r})\vec{\nabla}(\delta n(\mathbf{r},\omega)/n^2 K_T) + \vec{\nabla}\cdot\overset{\leftrightarrow}{\sigma}(\mathbf{r},\omega) \qquad (4.22)$$

and

$$-im\omega\delta\mathbf{v}_s(\mathbf{r},\omega) = -\vec{\nabla}[(n^2 K_T)^{-1}\delta n(\mathbf{r},\omega) - (Ts/c_V)\delta s(\mathbf{r},\omega)] + \vec{\nabla}\cdot\overset{\leftrightarrow}{\sigma}^{(s)}(\mathbf{r},\omega) \quad . \qquad (4.23)$$

The stress tensors entering eqns (4.22) and (4.23) are given by eqns (4.3) and (4.4), where the visco-elastic functions are those of the homogeneous superfluid at the local equilibrium densities, according to eqns (4.14) - (4.18), and can be calculated from the response functions of the homogeneous superfluid with the help of eqns (4.7) - (4.10). The thermodynamic coefficients entering the ALDA terms (the first term on the RHS of eqns (4.22) and (4.23)) are similarly given by the corresponding quantities for the homogeneous fluid at the local equilibrium density, as a direct consequence of the linearization effected in eqns (4.5) and (4.6) for the fluctuations of pressure and chemical potential. These ALDA terms are evidently responsible for the first and second sound modes in the superfluid.

## 5. Summary and discussion

In conclusion it will be useful to briefly summarize the main results that we have achieved in this work and to indicate some directions for future work.

The recent experiments performed on trapped vapours of alkali Bose atoms, measuring their collective excitations in wide ranges of frequency and temperature from the hydrodynamic regime to the collisionless regime, motivate the search for a unified theory of the dynamics of an inhomogeneous Bose superfluid.

In developing the basic time-dependent density functional framework for such a system in the linear response regime, we have focused our analysis on quantities having an immediate meaning, *i.e.* the condensate density, the superfluid velocity, and the current density. We have thus recognized that the essential building blocks of the response matrix of the superfluid derive from the two-by-two problem posed by the response of the density and phase of the condensate and from the irreducible part of the current response of the noncondensate. The difference between the condensate and the superfluid is evident from eqn (3.9), showing that the superfluid can be viewed as a condensate coupled to the noncondensate.

We have used the equations of motion for the density matrix and the condensate wave function to microscopically define the Kohn-Sham response functions of the condensate and the interaction kernels entering the whole linear response matrix of the superfluid. The basic equations that we have derived are eqn (2.30) for the condensate response and eqn (3.9) for the total current-current response, combined with eqn (B.1) for the irreducible part of the current-current response and eqn (B.7) for the condensate-irreducible current response. Two further results of our analysis are worth emphasizing: (i) the study of the dynamics of an inhomogeneous Bose superfluid at finite frequency requires attention to the coupling between the amplitude and the phase of the condensate, as related to the violation of the continuity equation for this component; and (ii) the microscopic, space- and time-dependent extensions of the density of normal fluid and superfluid emerge naturally from the theory (see eqn (3.6), (3.8) and (3.11)).

The above definition of the superfluid density has allowed us to introduce generalized hydrodynamic equations for a weakly inhomogeneous two-fluid model in isothermal conditions (see eqns (4.22) and (4.23)). We have shown that in this limit the visco-elastic functions are related through eqns (4.14) - (4.18) to the interaction kernels of the homogeneous fluid taken at the local equilibrium densities. We have also shown that local-density expressions hold for the thermodynamic coefficients responsible for first and second sound. As remarked in § 2.2, the fluctuations in condensate density decay on a faster time scale than the hydrodynamic fluctuations and accordingly we have not included this variable in the basic set for our generalization of hydrodynamic theory. This point may need further study. An extension of the theory to include couplings between density and temperature fluctuations may also be of some interest.

With regard to approximate calculations for a Bose superfluid, the visco-elastic spectra can be calculated from correlation functions of the homogeneous fluid at the local equilibrium density through eqns (4.7) - (4.10). In particular, the sound attenuation spectrum and the shear viscosity spectrum in eqns (4.7) and (4.8) can be calculated in a collisionless regime by means of the same two-pair decoupling scheme which has been used for an electron fluid in the normal state [22]. The results for these spectra as functions of temperature will be published elsewhere [37]. One recovers in this way for the sound-wave attenuation spectrum in a dilute Bose fluid at zero temperature the results previously obtained by Wong and Gould [25]. One can also show that in this approximation the value of $\eta(\omega)$ is simply proportional to that of $\zeta_2(\omega)$ [37].

More generally, a second-order perturbation expansion of the single-particle self-energies, combined with mode renormalization as in the so-called one-loop approximation introduced by Wong and Gould [25] provides a simple scheme allowing one to evaluate all the dissipation spectra in eqns (4.7) - (4.10) and the related mode-frequency shifts [38]. It turns out that within this approximation all these spectra can be expressed in terms of four exchange-correlation building blocks: a condensate kernel (the proper part of $\delta\alpha_{ex}/\delta n_c$), two noncondensate functions (the irreducible proper parts of the longitudinal and transverse current-current response) and a cross condensate-noncondensate term (the antisymmetric combination of the vertex functions $\Lambda_\alpha$). Actual calculation of these four functions yields that at all frequencies they take the same value aside from simple multiplying factors [38].

**Acknowledgments**

This work is supported by the Istituto Nazionale per la Fisica della Materia through the Advanced Research Project on BEC.

**Appendix A. Relations between excess kernels of the condensate and single-particle self-energies.**

We outline here an alternative derivation of eqn (2.36). We start from the expressions (2.12) - (2.15) for the response functions in terms of the weighed Green's functions introduced in eqn (2.10) and use the Dyson equations for the weighed Green's functions:

$$\overline{G}_{\alpha\beta} = \overline{G}^{(0)}_{\alpha\beta} + \sum_{\gamma,\delta} \overline{G}^{(0)}_{\alpha\gamma} \otimes \overline{\Sigma}_{\gamma\delta} \otimes \overline{G}_{\delta\beta} \quad , \tag{A.1}$$

where

$$\overline{\Sigma}_{\alpha\beta}(1,1') = [n_c(\mathbf{r}_1)n_c(\mathbf{r}_{1'})]^{-1} <\psi^\dagger_\alpha(\mathbf{r}_1)>_{eq} \Sigma_{\alpha\beta}(1,1') <\psi_\beta(\mathbf{r}_{1'})>_{eq} \tag{A.2}$$

are the weighed single-particle self-energies.

The equations of motions for the response functions $L$ are easily constructed from linear combinations of the set of equations (A.1), and the structure of eqn (2.36) is recovered with the following identifications:

$$\delta\vartheta_{ex} / \delta\varphi|_{n_c} = 2n_c(\mathbf{r}_1)n_c(\mathbf{r}_{1'})[\overline{S}^{(+)}_{11} - \overline{S}^{(+)}_{12} - \mu] \quad , \tag{A.3}$$

$$\delta\alpha_{ex} / \delta n_c|_\varphi = \frac{1}{2}[\overline{S}^{(+)}_{11} + \overline{S}^{(+)}_{12} - \mu] \quad , \tag{A.4}$$

$$\delta\vartheta_{ex} / \delta n_c|_\varphi = -in_c(\mathbf{r}_1)[\overline{S}^{(-)}_{11} - \overline{S}^{(-)}_{12}] \tag{A.5}$$

and

$$\delta\alpha_{ex} / \delta\varphi|_{n_c} = in_c(\mathbf{r}_{1'})[\overline{S}^{(-)}_{11} + \overline{S}^{(-)}_{12}] \quad , \tag{A.6}$$

where

$$\overline{S}^{(\pm)}_{11}(1,1') = \frac{1}{2}[\overline{\Sigma}_{11}(1,1') \pm \overline{\Sigma}_{22}(1,1')] \tag{A.7}$$

and

$$\overline{S}^{(\pm)}_{12}(1,1') = \frac{1}{2}[\overline{\Sigma}_{12}(1,1') \pm \overline{\Sigma}_{21}(1,1')] \quad . \tag{A.8}$$

Equations (A.3) - (A.8) make explicit the connection between the excess kernels and the combinations of symmetrized self-energies which were introduced in the homogeneous case by Wong and Gould [18]. Of course, for the homogeneous Bose fluid the evaluation of the ideal-gas response functions and the matrix inversion involved in eqn (2.37) are easily performed.

**Appendix B. Microscopic expressions for the current response**

We start from the definition (3.2) for the irreducible current response of the noncondensate at fixed condensate. From eqn (3.3) we obtain

$$\chi_{\tilde{j}\tilde{j}}(1,2) = \bar{\chi}_{\tilde{j}\tilde{j}}(1,2) + i(2m)^{-1}[\nabla_{1,1'}\tilde{G}_{1\gamma}(1,\bar{3})\tilde{G}_{\delta 1}(\bar{3}',1')]_{1=1'}$$
$$\times \frac{\delta \Sigma_{\gamma\delta}(\bar{3},\bar{3}')}{\delta \tilde{G}_{\varepsilon\eta}(\bar{4},\bar{4}')}\bigg|_{(n_c,v_s)} \chi_{\varepsilon\eta\tilde{j}}(\bar{4},\bar{4}',2) \quad , \quad (B.1)$$

where we have introduced the notations

$$\bar{\chi}_{\tilde{j}\tilde{j}}(1,2) = i(2m)^{-2}[\nabla_{1,1'}\nabla_{2,2'}\tilde{G}_{1\alpha}(1,2)\tilde{G}_{\alpha 1}(2',1')]_{1=1',2=2'} \quad (B.2)$$

and

$$\chi_{\alpha\beta\tilde{j}}(1,1',2) = (2m)^{-1}[\nabla_{2,2'}\frac{\delta \tilde{G}_{\alpha\beta}(1,1')}{\delta U(2,2')}]_{2=2'} \quad . \quad (B.3)$$

We remark that the function in eqn (B.3) reduces in the case $\alpha=\beta=1$ and $1'=1^+$ to the noncondensate density-current response function $\chi_{\tilde{n}\tilde{j}}(1,1')$. A Ward identity relating this latter function to $\chi_{\tilde{j}\tilde{j}}$ is derived in the main text (see eqn (3.5)).

We turn next to the off-diagonal (condensate-noncondensate) response functions. These can be expressed in terms of $\chi_{\tilde{j}\tilde{j}}$ and of the condensate response matrix introduced in § 2.2. In view of the symmetry property

$$\frac{\delta <\psi(1)>}{\delta U(2,2')}\bigg|_{\eta} = i\frac{\delta G_{11}(2,2')}{\delta \eta^*(1)}\bigg|_{U} \quad (B.4)$$

we discuss in detail only the dependence of the density and phase of the condensate on the potential $U$ at constant $\alpha$ and $\vartheta$. We proceed by functional differentiation of the equation of motion (2.14), after supplementing its LHS by the term $\int d\bar{2} U(1,\bar{2})<\psi(\bar{2})>$. We define a vertex function $\Lambda_\alpha(1,2,3)$ through

$$\frac{\delta <\psi_\alpha(1)>}{\delta U(2,2')}\bigg|_{\eta} = \tilde{G}_{\alpha\beta}(1,\bar{3})\Lambda_\beta(\bar{3},2,2') \quad . \quad (B.5)$$

This is related to the functional derivative entering eqn (3.6) by

$$\Lambda_\alpha(1,2,2') = \delta(1,2)<\psi_\alpha(\mathbf{r}_{2'})>_{eq} + [\frac{\delta \sigma_\alpha(1)}{\delta U(2,2')}]_{<\psi_\alpha>} \quad . \quad (B.6)$$

We then find for the condensate-noncondensate current response the result

$$\begin{pmatrix} \nabla_{3,3'}\frac{\delta n_c(1)}{\delta U(3,3')} \\ \nabla_{3,3'}\frac{\delta \varphi(1)}{\delta U(3,3')} \end{pmatrix}_{3=3'} = \begin{pmatrix} \chi_{n_c n_c}(1,\bar{2}) & \chi_{n_c\varphi}(1,\bar{2}) \\ \chi_{\varphi n_c}(1,\bar{2}) & \chi_{\varphi\varphi}(1,\bar{2}) \end{pmatrix}\begin{pmatrix} \overset{\mathbf{r}}{\alpha}_\Lambda(\bar{2},3) \\ \overset{\mathbf{r}}{\vartheta}_\Lambda(\bar{2},3) \end{pmatrix} \quad (B.7)$$

where

$$\overset{\text{r}}{\alpha}_{\Lambda}(1,2) = [n_{c,eq}(\mathbf{r}_1)]^{-1} \operatorname{Re}[\overset{\perp}{\Lambda}_1(1,2) < \psi^{\dagger}(\mathbf{r}_1) >_{eq}] \tag{B.8}$$

and

$$\overset{\perp}{\vartheta}_{\Lambda}(1,2) = 2 \operatorname{Im}[\overset{\perp}{\Lambda}_1(1,2) < \psi^{\dagger}(\mathbf{r}_1) >_{eq}] \quad , \tag{B.9}$$

with

$$\overset{\perp}{\Lambda}_1(1,2) = (2mi)^{-1} \nabla_{2,2'}[\Lambda_{\alpha}(1,2,2')]_{2=2'} \quad . \tag{B.10}$$

These equations are the natural extension of those given by Wong and Gould [25] for a homogeneous Bose fluid (see *e.g.* their eqns (2.26) and (2.27); see also eqn (5.8) in the book of Griffin [14]).

The vertex functions introduced above in the calculation of the off-diagonal response also enter to determine the reducible contribution to the current response. Starting from the definition

$$\chi_{\mathbf{jj}}(1,2) = i(2m)^{-2}[\nabla_{1,1'}\nabla_{2,2'} \frac{\delta G_{11}(1,1')}{\delta U(2,2')}|_{\eta}]_{1=1',2=2'} \tag{B.11}$$

where $G_{\alpha\beta}(1,1') = -i < T[\psi_{\alpha}(1)\psi_{\beta}^{\dagger}(1')] >$, we have

$$\frac{\delta G_{11}(1,1')}{\delta U(2,2')}|_{\eta} = \frac{\delta \tilde{G}_{11}(1,1')}{\delta U(2,2')}|_{(n_c,v_s)} + \frac{\delta G_{11}(1,1')}{\delta < \psi_{\alpha}(\overline{3}) >}|_{U} \frac{\delta < \psi_{\alpha}(\overline{3}) >}{\delta U(2,2')}|_{\eta} \quad , \tag{B.12}$$

the first term on the RHS being the irreducible part which is $\chi_{\tilde{\mathbf{jj}}}(1,2)$ in eqn (3.2). The reducible part in the RHS of eqn (B.12), from the definitions of the density and phase operators of the condensate in eqn (2.2), can be written in the form of a vector product, yielding

$$\chi_{\mathbf{jj}}^{(red)}(1,2) = (2m)^{-1} \left( \frac{\delta \mathbf{j}(1)}{\delta n_c(\overline{3})}|_{U} \quad \frac{\delta \mathbf{j}(1)}{\delta \varphi(\overline{3})}|_{U} \right) \begin{pmatrix} \nabla_{2,2'} \frac{\delta n_c(\overline{3})}{\delta U(2,2')}|_{\eta} \\ \nabla_{2,2'} \frac{\delta \varphi(\overline{3})}{\delta U(2,2')}|_{\eta} \end{pmatrix}_{2=2'} . \tag{B.13}$$

The second vector has been calculated in eqn (B.7). Using eqns (B.12) and (B.13) together with the symmetry property in eqn (B.4), we find the result

$$\chi_{\mathbf{jj}} = \chi_{\tilde{\mathbf{jj}}} + \left( \frac{\delta \mathbf{j}}{\delta v_s} \quad \frac{\delta \mathbf{j}}{\delta n_c} \right) \otimes \begin{pmatrix} \chi_{v_s v_s} & \chi_{v_s n_c} \\ \chi_{n_c v_s} & \chi_{n_c n_c} \end{pmatrix} \otimes \begin{pmatrix} \frac{\delta \mathbf{j}}{\delta v_s} \\ \frac{\delta \mathbf{j}}{\delta n_c} \end{pmatrix} \quad . \tag{B.14}$$

This is eqn (3.9) in the main text.

Finally, using eqns (B.7), (B.13) and (B.14) we obtain the results

$$\frac{\delta \mathbf{j}(1)}{\delta \varphi(2)}|_{U} = \overset{\text{r}}{\vartheta}_{\Lambda}(1,2) \tag{B.15}$$

and

$$\frac{\delta \mathbf{j}(1)}{\delta n_c(2)}\Big|_U = \overset{\leftrightarrow}{\alpha}_\Lambda(1,2) \quad . \tag{B.16}$$

These complete the evaluation of the current-current response.

As a final remark, we point out that in the homogeneous limit our eqns (B.10) and (B.14) reduce to those found by Griffin in the dielectric formalism (see ens (8.30) - (8.32) in ref. [14]). This immediately allows a diagrammatic interpretation of our results.

# References

[1]  D. S. Jin, J. R. Ensher, M. R. Matthews, C. E. Wieman, and E. A. Cornell, Phys. Rev. Lett. 77 (1996) 420.

[2]  M.-O. Mewes, M. R. Andrews, N. J. van Druten, D. M. Kurn, D. S. Durfee, C. G. Townsend, and W. Ketterle, Phys. Rev. Lett. 77 (1996) 988.

[3]  D. S. Jin, M. R. Matthews, J. R. Ensher, C. E. Wieman, and E. A. Cornell, Phys. Rev. Lett. 78 (1997) 764.

[4]  M. R. Andrews, D. M. Kurn, H.-J. Miesner, D. S. Durfee, C. G. Townsend, S. Inouye, and W. Ketterle, Phys. Rev. Lett. 79 (1997) 553 and 80 (1998) 2967.

[5]  D. M. Stamper-Kurn, H.-J. Miesner, S. Inouye, M. R. Andrews and W. Ketterle, preprint cond-mat/9801262 (1998).

[6]  A. Griffin, W.-C. Wu and S. Stringari, Phys. Rev. Lett. 78 (1997) 1838.

[7]  A. Minguzzi, M. L. Chiofalo and M. P. Tosi, Phys. Lett. A 236 (1997) 237.

[8]  E. Zaremba, A. Griffin and T. Nikuni, Phys. Rev. A 57 (1998) 4695.

[9]  G. M. Kavoulakis, C. J. Pethick and H. Smith, preprint cond-mat/9710130 (1997).

[10] V. B. Shenoy and T.-L. Ho, Phys. Rev. Lett. 80 (1998) 3895.

[11] T.-L. Ho and V. B. Shenoy, preprint cond-mat/9710275 (1997).

[12] P. C. Hohenberg and P. C. Martin, Ann. Phys. (NY) 34 (1965) 291.

[13] I. M. Khalatnikov, An Introduction to the Theory of Superfluidity (Benjamin, New York, 1965).

[14] A. Griffin, Excitations in a Bose-Condensed Liquid (University Press, Cambridge, 1993).

[15] E. Runge and E. K. U. Gross, Phys. Rev. Lett. 52 (1984) 997.

[16] O.-J. Wacker, R. Kümmel and E. K. U. Gross, Phys. Rev. Lett. 73 (1994) 2915.

[17] E. K. U. Gross and W. Kohn. Adv. Quantum Chem. 21 (1990) 255.

[18] E. K. U. Gross, J. F. Dobson and M. Petersilka, in Topics in Current Chemistry, edited by R. F. Nalewajski (Springer, Berlin, 1996), p. 1.

[19] G. Vignale and W. Kohn, Phys. Rev. Lett. 77 (1996) 2037.

[20] G. Vignale and W. Kohn, in Electronic Density Functional Theory, edited by J. Dobson, M. P. Das, and G. Vignale (Plenum, New York, 1997).

[21] H. M. Böhm, S. Conti and M. P. Tosi, J. Phys.: Condens. Matter 8 (1996) 781.

[22] S. Conti, R. Nifosí and M. P. Tosi, J. Phys.: Condens. Matter 9 (1997) L475.

[23] G. Vignale, C. A. Ullrich and S. Conti, Phys. Rev. Lett. 79 (1997) 4878.

[24] P. Szépfalusy and I. Kondor, Ann. Phys. (NY) 82 (1974) 1.

[25] K. Wong and H. Gould, Ann. Phys. (NY) 83 (1974) 252.

[26] J. Gavoret and P. Nozières, Ann. Phys. (NY) 28 (1964) 349.

[27] K. Huang and A. Klein, Ann. Phys. (NY) 30 (1964) 203.

[28] D. Forster, Hydrodynamic Fluctuations, Broken Symmetry, and Correlation Functions (Benjamin, Reading, 1975).

[29] A. Minguzzi and M. P. Tosi, J. Phys.: Condens. Matter 9 (1997) 10211.

[30] S. Giorgini, Phys. Rev. A 57 (1998) 2949.

[31] A. Minguzzi, S. Conti and M. P. Tosi, J. Phys.: Condens. Matter 9 (1997) L33.

[32] S. Stringari, Phys. Rev. Lett. 77 (1996) 2360.

[33] C. De Dominicis and P. C. Martin, J. Math. Phys. 5 (1964) 14 and 31.

[34] T. H. Cheung and A. Griffin, Phys. Rev. A 4 (1971) 237.

[35] J. W. Kane and L. P. Kadanoff, J. Math. Phys. 6 (1965) 1902.

[36] P. Nozières and D. Pines, The Theory of Quantum Liquids - Superfluid Bose Liquids (Addison-Wesley, Redwood City, 1990).

[37] A. Minguzzi, S. Conti and M. P. Tosi, to be published.

[38] M. P. Tosi, M. L. Chiofalo, A. Minguzzi and R. Nifosì, in New Approaches to Old and New Problems in Liquid State Theory, NATO-ASI Series, to be published.